# Critical pedagogy in the implementation of educational technologies

Parvathy Panicker
*Department of Computing*
*Goldsmiths University of London*
London, United Kingdom
ppani001@gold.ac.uk
ORCID:0000-0002-6934-6013

**Abstract**
This paper presents a critical review of the challenges to the implementation of learning technologies with particular focus on developing countries. A comprehensive literature review on learning technologies was undertaken for the purpose of understanding the challenges in developing countries. The research question is: what extent does education empower learners to be full participants in a socially democratic society? The literature review identified 25 papers relevant to this topic. Challenges are interrelated and to bring about changes in developing countries, this paper proposes two educational technology frameworks based on: 1. cultural conceptual framework, and 2. problem-based constructivist psychology simulation model. The framework and simulation model are both useful to guide practice and research.

**Introduction**

Learning technologies have gradually becoming implemented in developing counties and is believed to have a huge potential for governments struggling to meet the growing demand for education while facing a shortage of teachers (UNESCO, 2006). UNESCO states in the preamble of its Qingdao statement that equitable access to ICT skills and its relevant content is a priority for girls and women, persons with disabilities, internally displaced persons, socially or economically disadvantaged group, and other under-served populations (UNESCO, 2017). Students are not simply learners of an academic subject; they are social beings. The student population respond and react to social, political, cultural and organisational context around them.

Educational technology is "the study and ethical practice of facilitating learning and improving performance by creating, using and managing appropriate technological processes and resources"( Bradshaw AC 2017,p 12). In 2008, Association for Educational Communications Technology (AECT) reviewed their definition of educational technology to include ethics in order to keep culture-related issues such as inclusion foreground in work. Furthermore, the definition acknowledges that individuals can belong to multiple cultures, sharing the characteristics of each culture to a certain degree (Bradshaw AC 2017, p3). Individuals and educational technology professional organisations are increasingly recognising the urgent need to understand complex interactions between culture, learning and technology. Because e-learning most often is transferred from the developed world, there is a need to understand what challenges are already experienced and to what extent these are met in developed countries. There is also a need to understand which additional challenges, if any, there may be in developing countries (Andersson, 2009). E-learning technology, which is a cheaper and more flexible alternative can be seen as a tool for students who have access to higher education. Challenges are however plentiful. In many developing countries there is lack of proper curriculum is non-existent, funds are lacking, and privacy-security are not being appropriately addressed (Qureshi ,2012). Resistance to change, limited expertise, computer literacy, need for face-to-face interaction, English competency and level of awareness were some of the challenges (Qureshi, 2012; Oye ND , 2011).

**Implementation of educational technology in different countries**
When culture bump is used as a pedagogical strategy and the skills are taught as a means of communication then transformative education is likely to occur (Archer, 2012). The transformative education empowers socially marginalized people thereby enabling a societal shift toward social justice (Bradshaw AC 2017, p10). The transformative education allows an individual to ask questions and seek answers to it in the 'problem posing' model compared to the 'banking' model of transmissive education. Transformative education requires and supports higher order cognition and allows a flexible open-ended way of exploration to seek questioning and discovery. In the primary school teachers' acceptance of educational technology in China have identified factors such as facilitating conditions and attitude as most significant (Wong GK, 2016).

It has been observed that teachers are generally busy with administrative duties, and to manage the heavy workload, teachers generally believe that the most efficient method to prepare for and deliver a lesson is through didactic approach using 'chalk and talk' resources. They take up educational technologies at the insistence of senior management to enhance their teaching effectiveness. Teachers consider technology-aided education as a supplementary to work performance rather than as a requirement. Women in Namibia and Tanzania have recognised importance of developing post literacy skills and have used mobile phone apps to apply literacy skills learned in school to support their entrepreneurial work or political activism (Semali, 2014). Digital literacy enhanced their personal communications and strengthened social and business networks, inter-group relations and linguistic skills. It was also observed that new words were added to the vocabulary overlapping with existing Kiswahili terms e.g., *intaneti*(internet), *salio*(balance) in the everyday lexicon in Tanzania. It is therefore evident that the challenges faced by developing countries to the implementation of leaning technologies is unique and cannot be compared to the challenges faced by developed countries.

In India, the traditional '*gurukula*' embedded the essence of oral culture and knowledge was transferred over several generations. Invasions from different powers such as the Mughals and the British led to a gradual dismantling of a rather successful model for education. A new system of education was thrust upon the Indian population and at the time of independence in 1947 the British system of education was adopted and was retained. This system was not culturally adapted and many feel that this has led to the stunting of development and progress in the fields of science, technology and mathematics. The colonial layout of infrastructure had everlasting effect on education and technology use. The unequal distribution of wealth combined with communication infrastructure coupled with other factors such as the markets costs to users have had its significant impact on the technological and informational diffusion.

Furthermore, technologies such as distance education, e-learning and educational technologies originated in the West, it would be irrelevant to compare its implications to an oral culture that exists in most developing countries. By identifying the challenges this review paper pays special attention to progressive education through critical pedagogy. The factors that have been identified in this paper are the most significant from an anthropological point of view when considering world cultures. The overarching goal with this article is to provide critical reflections to encompass educational approaches in order to empower learners to be full participants in post-colonial democratic societies.

**Post-colonialism**
With increasing globalisation and immigration, and a tendency to copy the western world, it becomes difficult if not impossible to essentialise national and cultural identities. The quintessential postcolonial individual might be part of two or more cultures but not feeling as belonging to either, because he/she is transnational (Ha, M.P., 1997). Overall, theories about culture identities provide rich insights for researchers, practitioners and students of culture and communication. Together these theories offer rich understandings of cultural identity regarding how cultural identities are communicated and being applied in life (Ha, M.P., 1997). Some authors argue that the decolonisation of the post-war decades did not lead to the true independence and emancipation of poor countries, since they were tied up with a global capitalist system where they were bound to lose. At the level of culture, writers influenced by neo-imperialism argued that formally colonised people became more dependent on the models and knowledge systems of the former colonisers.

**Economy and religion**
In a capitalist economy, money is the common denominator for economic activity. In a capitalistic country market economy can be thought of as going to supermarket to buy; consumer might use any one of the leading supermarkets one time and may use an entirely new one the next time he/she shops. In other words, a capitalist economy recognises only one form of commodity exchange, namely market exchange based on laws of supply and demand. In the gift economy, on the other hand, the distribution of goods takes place with no fixed price. In societies where the exchange of gifts is widespread, this contributes significantly to systemic integration (Eriksen, 2015). For example, 'guru-shishya' tradition or lineage was observed between teachers and disciples in Hindu, Jain, Buddhist and Sikh religion. In this educational culture the student eventually masters the knowledge that the 'guru' or teacher embodies, by staying with the guru at his home and in exchange the 'shishya' or the student offers 'gurudakshina' as a gift to the guru as a token of gratitude. Such token can be a simple fruit, flowers or even a part of one's own body.

**Motivational Grit for sustenance**
Successful students often taste failure, but they also know that perseverance and constant effort lead to their goals (Aparicio M, 2016). The non-cognitive trait grit has been introduced recently in studies (Duckworth A,2007). Grit is defined as the 'perseverance and passion for long-term goals'. Grit was studied along with other non-cognitive measures such as individual stamina, perseverance. Grit is different when compared to self-control and will-power. According to literature, cognitive measures include IQ tests, aptitude tests. Grit did not relate to IQ, but rather is highly correlated with the Big Five Conscientiousness (Duckworth A,2007). One important determinant of success is grit- the tenacious pursuit of a dominant superordinate goal despite setbacks (Duckworth A, 2014). Grit entails working assiduously towards a single challenging goal through thick and thin, on a timescale of years or even decades (Duckworth A, 2014). Grit in learning can be illustrated as a turtle in a race, and compared with the hare, has persistence and is goal directed. In contrast to the turtle, the hare only runs quickly and seeks short-term results (Aparicio M, 2017). In other words, students who graduated on schedule were grittier, and grit was a more powerful predictor of graduation (Duckworth A, 2016).
Older individuals tend to be higher in grit than younger individuals, suggesting that the quality of grit, although a stable individual difference, may nevertheless increase over their life span. Grittier individuals made fewer career changes than less gritty peers of the same age (Duckworth A, 2007). However, rare studies have examined the effects of national culture and grit for knowledge transfer in a virtual environment. Cultural grit has never been measured. Refer to table summarises the cultural and individual factors for implementation of educational technologies in different cultures.

Conceptual framework for implementation of educational technologies

| Cultural | Individual factors |
| --- | --- |
| Post colonial theory | Grit-passion |

| Societal history | Grit-perseverance |
| Economy and world religion | |

**Theoretical frameworks: critical pedagogy**
Critical pedagogy is a framework that encourages critical theory. This theory allows researchers to argue and think critically and mainly focusses on social inclusion of all. It is primarily concerned with individuals or communities that have been excluded as a result of their rights being taken away from them. According to anthropology, some of the groups from whom their rights have been taken away are: women, people in prison, senior citizens, people with disability and sometimes vulnerable young people. Critical theory tries to answer questions such as who is included or excluded, and whose interests are represented, excluded or marginalised. Grounded in concerns for social justice, fairness, and equality, critical pedagogy brings together critical theory, liberation ethics, and progressive education (Bradshaw AC 2017, p12). With the right understanding of criticality, it is a commitment to dig deeper in order to understand the roots and ramifications to allow full participation and humanity of all. Therefore, criticality requires awareness to the struggles of those at the margins of society, self-interrogation, self-examination, learning regarding one's own positionalities (Bradshaw AC 2017, p12). Education for critical consciousness is a socio-political educative approach that engages learners in questioning the nature of their historical and social situation which Freire described as "reading the world". Developing critical perception is an ongoing process that cannot be imposed. By developing critical consciousness, individuals are harder to control, and people will resist the dominant culture's actions that go against their own interest. Individuals often accommodate power structures and hierarchy and we are made to believe someone is in charge always. In such structures, individuals are sometimes not mindful about the harm that has been caused to others as a result of the benefits received and may continue to maintain blind and in denial.

Critical pedagogy is a commitment to freedom in multiple forms- freedom from oppressive practices and limiting perspectives and from mythologies and manipulations that facilitate oppression and hegemony (Bradshaw AC 2017, p12). Open data and Open educational resources when supported by a critical pedagogy scaffolding, would represent a powerful resource for learners in the context of society affected by trauma. The enormous potential of open data in education resides in the fact that their use can trigger critical reflections for empathic care: the act of discovering, researching, getting to know about reality through open data is a modality of care that refers to pedagogic action (Manca A, 2017). The answers to questions about integrating technologies into education lies not at the level of technical skill but at the more fundamental level of assumptions. Educators must be allowed to recognise, reclaim and encourage to critically reflect upon their practices with technology.

There is a need to develop educational technology implementation strategies for societies that have been affected by trauma based on the critical pedagogy theory outlined above. Hence the following recommendations are proposed. Based on theoretical considerations, two different frameworks for implementing educational technology are proposed:

**Proposal #1: Participatory action**
Participatory action refers to "a cycle of inquiry involving plan-act-observe-reflect" (Herr & Anderson, 2005). This approach incorporates the collaboration of local community members to bring change in a society. These approaches have the potential to be psycho-politically valid by producing action and knowledge that contribute to social change: they offer researchers and community members the opportunity "to 'research back,' in the same tradition of 'writing back' or 'talking back' that characterizes much of the postcolonial or anticolonial literature". Participatory development interventions such as poetry and photography for teens have proved to be suitable as a model of psychotherapy that empowers and leads towards social change. For e.g., in Czech Republic participatory action has been used to address and overcome deep rooted obstacles of educational system. Thus, participatory action is seen as a 'sustainability science' which strives for social and scientific impact. Participatory action approaches represent an opportunity for mental health professionals to address questions from a social and methodological stance that is less likely to merely reproduce new forms of colonized theory and practice (Lykes & Coquillon, 2007). Fields such as public health and education have established traditions of participatory action (Minkler, M. and Wallerstein, 1997), yet these approaches are largely unfamiliar and little-utilized across educational technologies.

Through participatory action, students will have a voice about the courses, type of courses, and the method of course layout that they intend on registering. One of the most effective ways in which this can move toward as a culturally inclusive educational technology is though allowing the students to participate in discussions and recommending courses that they think might be most suitable for them. In this way the learners are in control of their learning and can approve or disapprove courses.

**Proposal #2: Constructivist psychology- simulation model**
Another approach is to build a problem-based model that empowers learning through a transformative model rather than a banking system of transmission. An adaptive media uses a simulation model that provides intrinsic feedback to students for a society where oppression itself is the pathogen (Smith L, 2009). An equilibrium between an adaptive medium of continual activity and a narrative medium of continual receptivity should be introduced. Along with this, short clips of video to keep the attention span of students should also be considered. Minimal text should be provided along with the video in order to engage learners to avoid boredom. In this way, the students fall into neither sit-forward or sit-back learning and their attention span is maintained. In this system, the learner is monitored continuously by the system and feedback is provided as recommended topics or courses that could benefit them in the future. The system should also include tutorial programmes as it is premised on the idea that it is possible for a computer program to emulate a teacher (Laurillard, D., 2002). In this way extrinsic feedback could also be given for the learner.

**Conclusion**
This paper presents a critical review of the challenges to the implementation of learning technologies. Challenges are interrelated and to bring about changes in developing countries, this paper proposes two educational technology frameworks based on: 1. cultural conceptual framework, and 2. problem-based constructivist psychology simulation model. The framework and simulation model are both useful to guide practice and research.